\documentclass{ptephy_v1}

\usepackage{boites}
\usepackage{amsmath,amssymb}
\usepackage{graphicx}
\usepackage{float}
\usepackage{tabularx}
\usepackage{xcolor}

\def\<{\left\langle}
\def\>{\right\rangle}
\def\({\left(}
\def\){\right)}
\def\e{{\rm e}}
\def\Det{{\rm Det\,}}
\def\dps{\displaystyle}
\newcommand\bs[1]{\boldsymbol{\mathit{#1}}}

\begin{document}

\title{Tracy--Widom method for J\'{a}nossy density and joint distribution of extremal eigenvalues of random matrices}

\author{\name{\fname{Shinsuke M.} \surname{Nishigaki}}{\ast}} 

\address{\affil{~}{Graduate School of Natural Science and Engineering, Shimane University, Matsue 690-8504, Japan}
\email{mochizuki@riko.shimane-u.ac.jp}}

\begin{abstract}%
The J\'{a}nossy density for a determinantal point process is the probability density that an interval $I$
contains exactly $p$ points except for those at $k$ designated loci.
The J\'{a}nossy density associated with an integrable kernel 
$\mathbf{K}\doteq (\varphi(x)\psi(y)-\psi(x)\varphi(y))/(x-y)$
is shown to be expressed as a Fredholm determinant 
$\mathrm{Det}(\mathbb{I}-\tilde{\mathbf{K}}|_I)$ of a transformed  kernel 
$\tilde{\mathbf{K}}\doteq (\tilde{\varphi}(x)\tilde{\psi}(y)-\tilde{\psi}(x)\tilde{\varphi}(y))/(x-y)$.
We observe that $\tilde{\mathbf{K}}$ satisfies Tracy and Widom's criteria if $\mathbf{K}$ does,
because of the structure that the map $(\varphi, \psi)\mapsto (\tilde{\varphi}, \tilde{\psi})$
is a meromorphic
$\mathrm{SL}(2,\mathbb{R})$ gauge transformation between covariantly constant sections.
This observation enables application of the Tracy--Widom method to 
J\'{a}nossy densities, expressed in terms of a solution to a system of differential equations in the endpoints of the interval.
Our approach does not explicitly refer to isomonodromic systems associated with Painlev\'{e} equations
employed in the preceding works.
As illustrative examples we compute J\'{a}nossy densities with $k=1, p=0$ for Airy and Bessel kernels,
related to the joint distributions of the two largest eigenvalues of random Hermitian matrices
and of the two smallest singular values of random complex matrices.
\end{abstract}

\subjectindex{A10, A13, A32, B83, B86}

\maketitle

\section{Introduction}
\begin{figure}[h] 
\begin{center}
\hspace{1cm}
\includegraphics[bb=0 0 360 237,width=75mm]{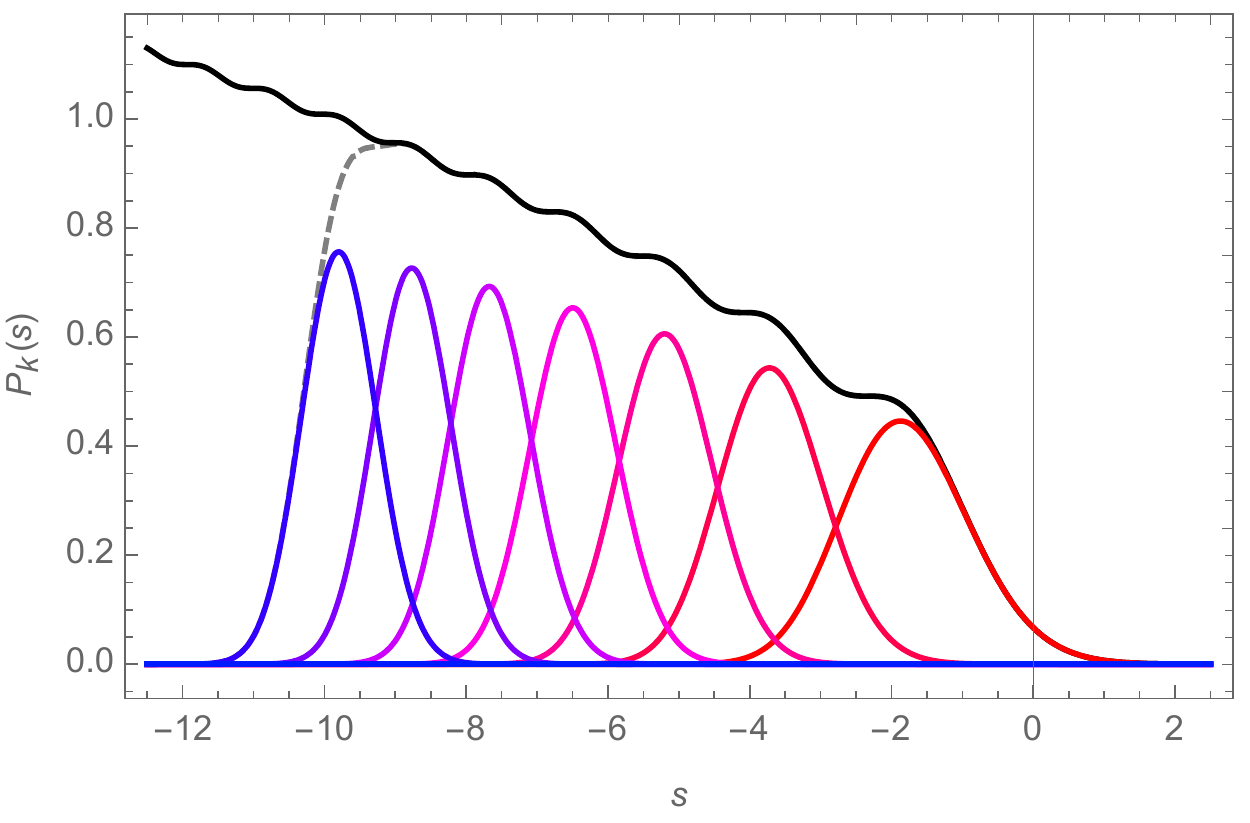}~~~
\includegraphics[bb=0 0 360 235,width=75mm]{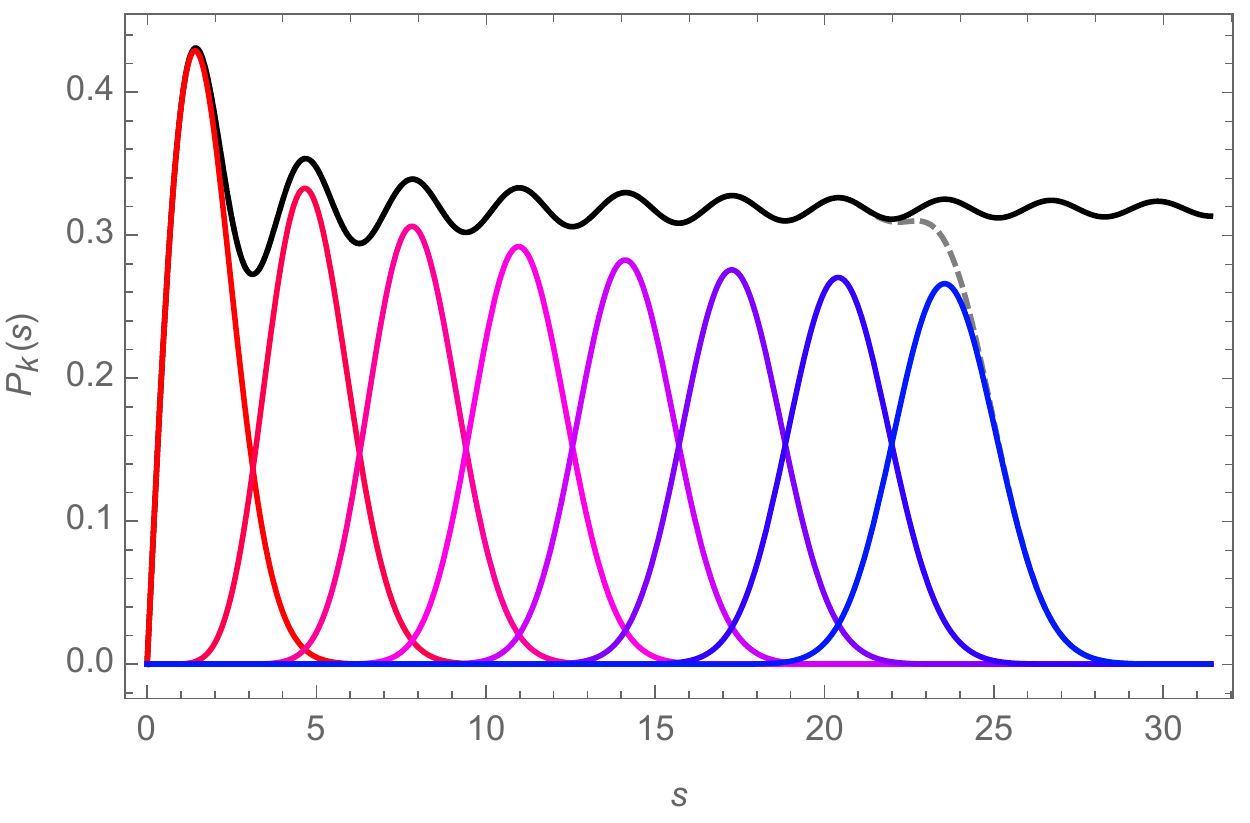}
\caption{\label{fig:Pk}
Distributions $P_k(s)$ of the scaled $k$th largest eigenvalues of random Hermitian matrices
(Tracy--Widom distribution) (right)
and of the $k$th smallest singular values of random complex square matrices (left)
(red ($k=1$) to blue ($k=8$)),
their sums $\sum_{k=1}^8 P_k(s)$ (grey dotted), and the spectral densities $\rho_1(s)$ (black).
}
\end{center}
\end{figure}

In the history of random matrix theory (RMT), which models the local fluctuation of energy
levels of quantum chaotic and/or disordered Hamiltonians typified by the Sinai billiard, the 
Anderson tight-binding model and the QCD Dirac operator, Gaudin and Mehta's discovery
that the distribution of ordered eigenvalues or their spacings is expressed in terms of 
the Fredholm determinant or Pfaffian of an integral kernel $\mathbf{K}_I$ restricted on an interval $I$ \cite{Mehta,Forrester}
has been known as long as the RMT itself.
Specifically, the distribution $P_k(s)$ of the $k$th largest eigenvalue (centered and scaled) 
of random Hermitian matrices is given as
\begin{align}
P_k(s)=
\frac{1}{k!}\partial_s\left.\left(-\partial_{{z}}\right)^k
\mathrm{Det}\bigl(\mathbb{I}-{z}{\mathbf{K}}_{(s,\infty)}\bigr)\right|_{{z}=1}
\label{PkDet}
\end{align}
with $\mathbf{K}$ being the Airy kernel \cite{Tracy:1994a},
and that of the $k$th smallest singular values of random complex matrices
is given by Eq.~(\ref{PkDet}) with $\mathbf{K}$ being  the Bessel kernel \cite{Tracy:1994b}
(with replacements $I=(s,\infty)\mapsto (0,s)$ and $\partial_s\mapsto -\partial_s$).
These trains of peaks that gradually approach Gaussian in the spectral bulk 
\cite{Gustavsson:2005}
constitute the spectral densities $\rho_1(s)=\sum_{k=1}^\infty P_k(s)$,
as plotted in Fig.~\ref{fig:Pk}.
For the practical purpose of fitting some spectral data to the RMT to extract system-specific constants
(such as the chiral condensate and the pion decay constant in the case of 
QCD Dirac operators \cite{Yamamoto:2018}),
characteristic peaky shapes of the individual distributions are better suited than the spectral density,
as the oscillation of the latter tends to smooth out in the bulk and the data-fitting
would yield little more than the mean level density.

\begin{figure}[b] 
\begin{center}
\hspace{1cm}
\includegraphics[bb=0 0 360 243,width=78mm]{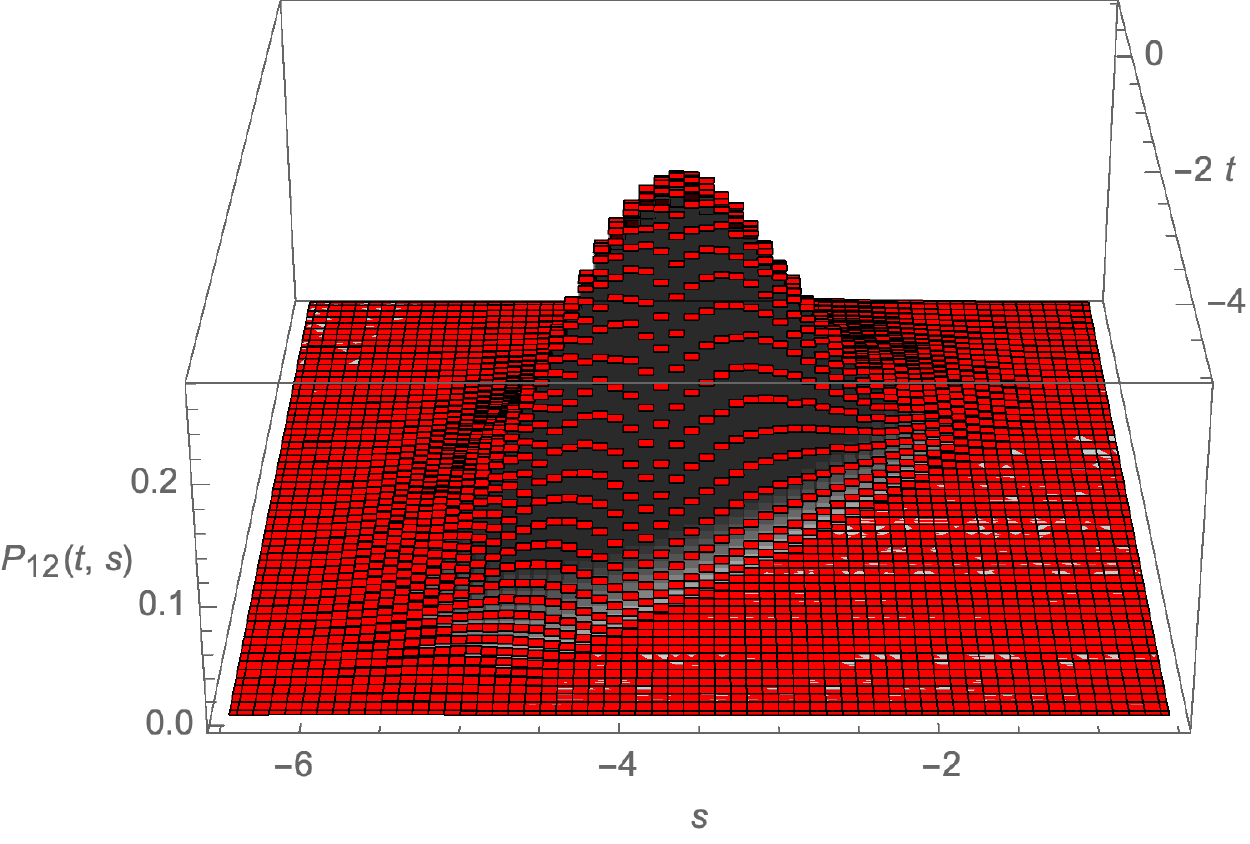}~
\includegraphics[bb=0 0 360 256,width=74.5mm]{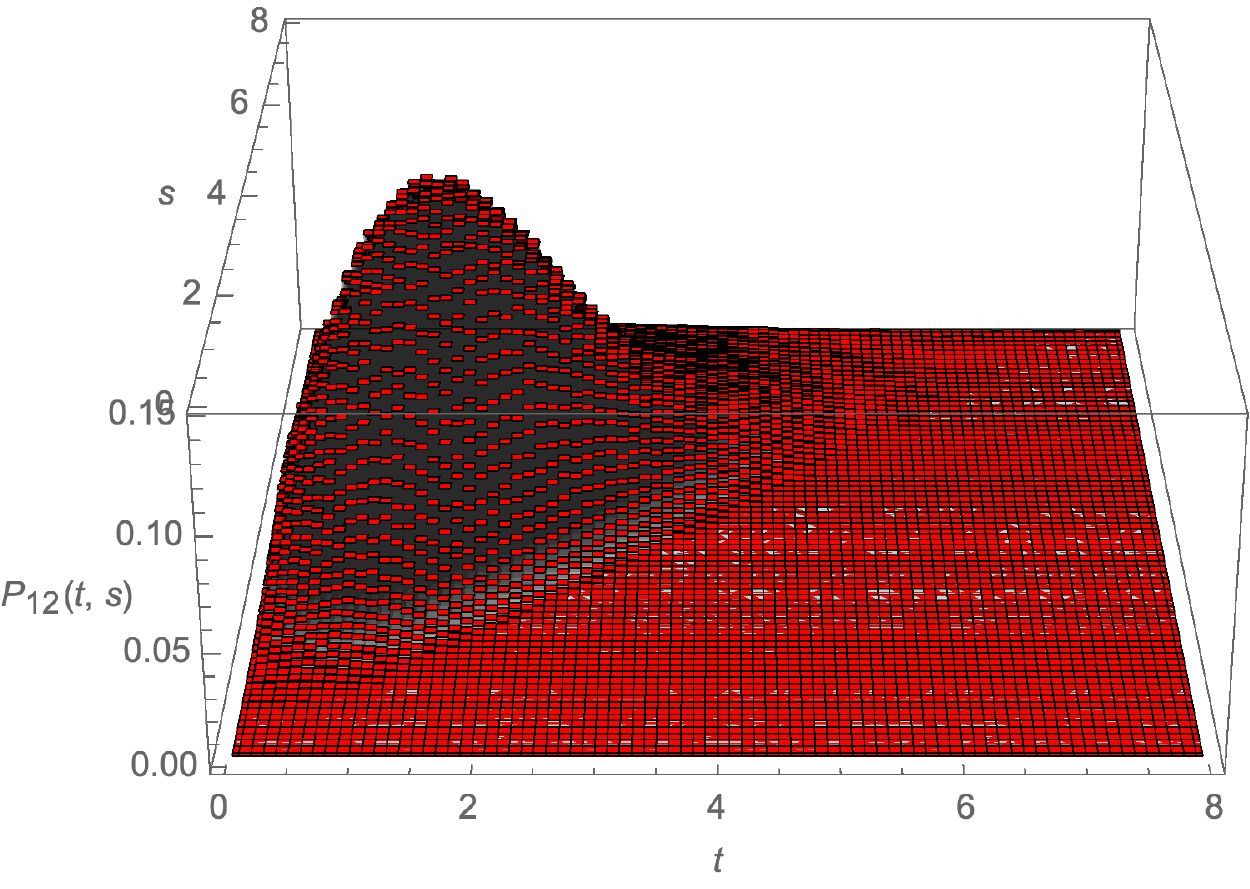}
\caption{\label{histograms}
Histograms
of the first $(t)$ and second $(s)$ largest eigenvalues of random Hermitian matrices (left)
and of the first  $(t)$ and second  $(s)$ smallest singular values of random complex square matrices (right).
Matrix rank $N=128$ and number of samples $=10^7$ for each case, and  eigen/singular values $x$ are rescaled as:
$t$ or $s=\sqrt{2}N^{1/6}(x-\sqrt{2N})$ and  $\sqrt{2N}x$, respectively.
}
\end{center}
\end{figure}

With this in view, the purpose of this article is to advance the formula (\ref{PkDet}) a step further 
and provide a ``user-friendly'' analytic method to compute
the joint distribution $P_{1\cdots k}(s_1,\ldots,s_k)$ of the first to $k$th 
largest/smallest eigenvalues of unitary-invariant random matrices,
which is a constituent of the $k$-point correlation function $\rho_k(s_1,\ldots,s_k)$.
To this end we apply the strategy of Tracy and Widom \cite{Tracy:1994c}
on the evaluation of Fredholm determinants of integrable integral kernels to the J\'{a}nossy density
\cite{Daley,Borodin:2000,Borodin:2003,Soshnikov:2003}, 
i.e.,~the probability distribution that an interval contains no eigenvalue except for those at $k$ designated loci.
As the simplest examples we shall evaluate
the joint distributions $P_{12}(t,s)$ of the first and second largest eigenvalues and smallest singular values
(see Fig.~\ref{histograms} for their histograms),
i.e.,~the first peak that constitutes the two-point correlation function
$\rho_2(t,s)=\sum_{k<\ell}P_{k\ell}(t,s)$, for the Airy and Bessel kernels.
Each case has been worked out previously in Refs.~\cite{Forrester:2007,Witte:2013},
which devised an elaborate analytic procedure
involving the Painlev\'{e} II and III${}'$ transcendents and the associated isomonodromic systems \cite{Jimbo:1980}.
This approach is later simplified (for the Airy kernel)  
using a solution to the Lax pair associated with the Painlev\'{e} XXXIV system \cite{Perret:2014}.
Our alternative method presented in this article, which does not explicitly refer to these systems and
employs the familiar Tracy--Widom method,
has a clear advantage of permitting straightforward generalizations to a general $P_{1\cdots k}$
and/or to various finite-$N$ and large-$N$ kernels 
(Hermite, Laguerre, and other hypergeometric; circular, beyond-Airy, $q$-orthogonal, etc.)
appearing in the RMT.

This article is composed of the following parts: In Sect.~2 we list known formulas on 
J\'{a}nossy densities of a determinantal point process, and then
express them in terms of Fredholm determinants of the ``transformed kernel" $\tilde{\mathbf{K}}$.
The latter is a novel presentation to the best of our knowledge,
except for the simplest ($k=1$) case of the sine kernel previously treated in Ref.~\cite{Forrester:1996}.
In Sect.~3 we demonstrate that $\tilde{\mathbf{K}}$ satisfies Tracy and Widom's criteria
for their functional-analytic method to be applicable if the original $\mathbf{K}$ does.
In Sect.~4 we evaluate J\'{a}nossy densities and joint distributions of 
the first and second extremal eigenvalues from the Airy and Bessel kernels by the Tracy--Widom method.
In Sect.~5 we conclude with listing possible applications and extensions of our approach.
Numerical data of J\'{a}nossy densities for the Airy kernel and
the Bessel kernels at $\nu=0, 1$ are attached as supplementary material.
Throughout this article we follow the notations of Ref.~\cite{Tracy:1994c}, hereafter denoted as TW.

\section{J\'{a}nossy density}
First we collect some facts on determinantal point processes (DPPs).
Let $\mathfrak{X}$ be a countable set.
Consider an ensemble of finite subsets of $\mathfrak{X}$ consisting of $N$ elements (``particles") $(n_1,\ldots,n_N)$,
and assign to them a joint probability in a determinantal form:
\begin{align}
P(n_1,\ldots,n_N)=\frac{1}{N!}\det\left[K(n_i,n_j)\right]_{i,j=1}^N,\quad n_i\in\mathfrak{X}.
\end{align}
Here $\mathbf{K}=\left[K(n,m)\right]_{n,m\in\mathfrak{X}}$ is an operator in the Hilbert space $L^2(\mathfrak{X})$,
i.e.,~an infinite-dimensional matrix indexed by the points of $\mathfrak{X}$.
The operator
$\mathbf{K}$,
which we also call a kernel, is required to be real, symmetric, projective, and normalized:
\begin{align}
\mathbf{K}=\mathbf{K}^*=\mathbf{K}^t,\quad
\mathbf{K}\cdot \mathbf{K}=\mathbf{K},\quad \mathrm{tr}\,\mathbf{K}=N.
\label{projnorm}
\end{align}
These requirements lead to the $k$-point correlation function $\rho_k(n_1,\ldots,n_k)$,
i.e.,~the joint probability that $k$ particles occupy the points $n_1,\ldots,n_k$,
to be given in a determinantal form as well:
\begin{align}
\rho_k(n_1,\ldots,n_k)=\det\left[K(n_i,n_j)\right]_{i,j=1}^k:=\det {\boldsymbol\kappa} .
\label{Rk}
\end{align}

\begin{figure}[h] 
\begin{center}
\vspace{-8mm}
 \includegraphics[bb=0 0 360 223,width=105mm]{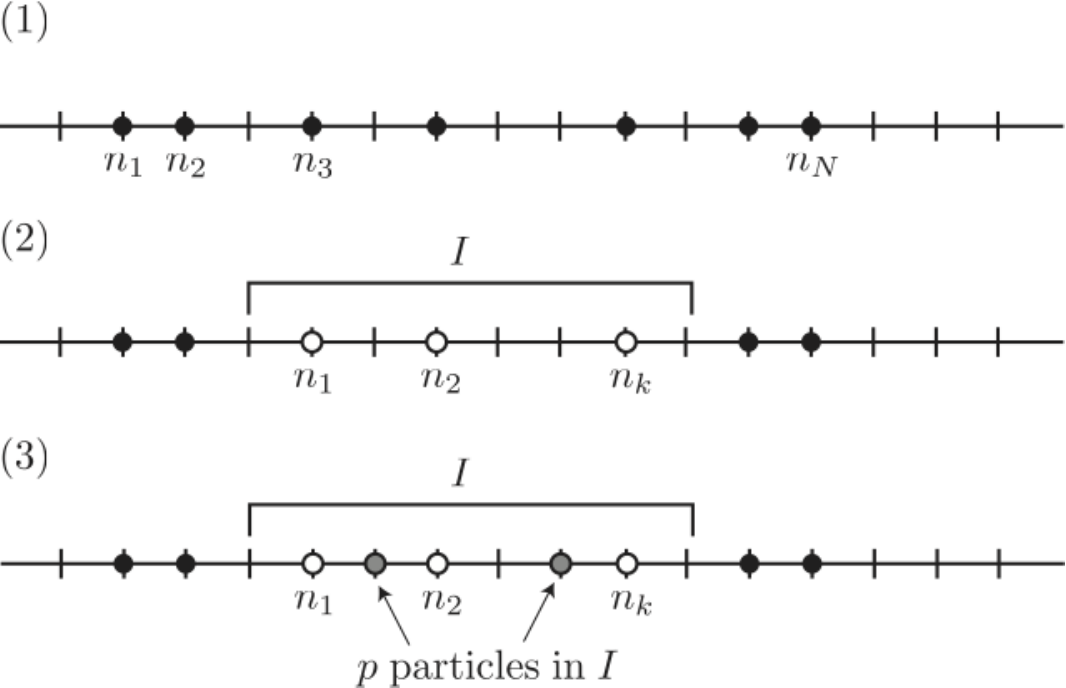}
\caption{\label{fig:Janossy}
Distribution of particles in a DPP. 
(1) $N$ particles distributed exclusively on $N$ loci $n_1,\ldots,n_N$ in $\mathfrak{X}$.
(2) Exactly $k$ particles in ${I}$, one at each of the $k$ designated loci $n_1,\ldots, n_k$ in ${I}$.
(3) $k$ particles, one at each of the $k$ designated loci $n_1,\ldots, n_k$
and other exactly $p$ particles on $p$ undesignated loci in ${I}$.}
\end{center}
\end{figure}

The J\'{a}nossy density $J_{k}(n_1,\ldots,n_k;{I})$ 
is defined as a probability that there is no particle in a subset
${I}\subset \mathfrak{X}$ except for $k$ particles, one at each of the $k$ designated loci $n_1,\ldots,n_k$
 (see part (2) in Fig.~\ref{fig:Janossy}).
The restriction that $n_k\in I$ could actually be lifted.
Using Eq.~(\ref{Rk}),
the J\'{a}nossy density is given in terms of the  kernel restricted on ${I}$,
$\mathbf{K}_{I}=\left[K(n,m)\right]_{n,m\in {I}}$
 (see, e.g.,~page 341 of Ref.~\cite{Borodin:2000}, where it is denoted as $\pi(X)$):
\begin{align}
J_{k}(n_1,\ldots,n_k;{I})&=\det(\mathbb{I}-\mathbf{K}_{I})\cdot \det\left[
\langle n_i|\mathbf{K}_{I}(\mathbb{I}-\mathbf{K}_{I})^{-1}|n_j\rangle
\right]_{i,j=1}^k.
\label{Janossy_prob0}
\end{align}
We interchangeably use notations $A(n,m)=\<n|\mathbf{A}|m\>$ for the $(n,m)$-element of a matrix $\mathbf{A}$.
Now, making use of an identity
$\det{ \left|\begin{array}{cc} {\mathbf A} & {\mathbf B} \\ {\mathbf C} & {\mathbf D}\end{array}\right|}
=\det {\mathbf D}  \cdot \det\left({\mathbf A}-{\mathbf C}{\mathbf D}^{-1}{\mathbf B}\right)$ repeatedly,
we present two alternative expressions for the J\'{a}nossy density \cite{Fuji:2019}:
\begin{align}
J_{k}(n_1,\ldots,n_k;{I})&
=(-1)^k\det
\left|
\begin{array}{ll}
\left[-K(n_i,n_j)\right]_{i,j=1,\ldots,k} &
\left[-K(n,n_j)\right]_{n\in {I};\;j=1,\ldots,k}\\
\left[-K(n_i,m)\right]_{i=1,\ldots,k;\;m\in{I}}&
\left[\delta_{nm}-K(n,m)\right]_{n,m\in {I}} 
\end{array}
\right|
\nonumber\\
&:=
(-1)^k\det
\left|
\begin{array}{cc}
-{\boldsymbol\kappa} & -\bs{k} \\
-\bs{k}^t & \mathbb{I}-\mathbf{K}_I
\end{array}
\right| 
\nonumber\\
&=
\det{\boldsymbol\kappa}\cdot
\det\(\mathbb{I}-\mathbf{K}_I+(\bs{k}^t {\boldsymbol\kappa}^{-1} \bs{k})_I \).
\label{Janossy_prob2}
\end{align}
Equation (\ref{Janossy_prob2}) indicates that 
{\em the J\'{a}nossy density is a gap probability for a DPP with a transformed kernel}
\begin{align}
\tilde{\mathbf{K}}=\mathbf{K}-\bs{k}^t {\boldsymbol\kappa}^{-1} \bs{k},
\label{Ktilde}
\end{align}
multiplied by the $k$-point correlation function $\det {\boldsymbol\kappa}$.
The projectivity and the normalization conditions (\ref{projnorm}) with $N\mapsto N-k$ 
can be verified for $\tilde{\mathbf{K}}$
in a straightforward manner.
Note that
\begin{align}
&\tilde{\rho}_p(m_1,\ldots,m_p; n_1,\ldots,n_k):=
\frac{\rho_{p+k}(m_1,\ldots,m_p, n_1,\ldots,n_k)}{\rho_{k}(n_1,\ldots,n_k)}
=\det\left[\tilde{K}(m_i,m_j)\right]_{i,j=1}^p
\label{tilderhop}
\end{align}
represents the {\em conditional} joint probability that $p$ particles occupy the points $m_1,\ldots,m_p$
under the presumption that $k$ particles already occupy the points $n_1,\ldots,n_k$.
Obviously, this fact could as well be deduced from the very definition of the conditional joint probabilities;
e.g.,~for $k=1$, the transformed kernel ${\displaystyle \tilde{K}(m,m')=K(m,m')-\frac{K(m,n)K(n,m')}{K(n,n)}}$ satisfies
\begin{align}
\tilde{\rho}_1(m; n)&=
\frac{\rho_{2}(m,n)}{\rho_{1}(n)}
=\frac{K(m,m)K(n,n)-K(m,n)K(n,m)}{K(n,n)}
=\tilde{K}(m,m),
\nonumber\\
\tilde{\rho}_2(m_1,m_2; n)&=
\frac{\rho_{3}(m_1,m_2,n)}{\rho_{1}(n)}
\\
&=\frac{K(m_1,m_1)K(m_2,m_2)K(n,n)\pm (\mbox{5 terms})}{K(n,n)}
=\det\left[\tilde{K}(m_i,m_j)\right]_{i,j=1}^2,
\nonumber
\end{align}
etc.,
without any recourse to J\'{a}nossy densities.
We could have reversed the course of derivation of J\'{a}nossy densities (\ref{Janossy_prob2})
and started backward from Eq.~(\ref{tilderhop}).

Generalization to the probability $J_{k,p}(n_1,\ldots,n_k;{I})$ that there are exactly $p$ particles in ${I}$ 
except for $k$ particles, one at each of the $k$ designated loci, is straightforward (see part (3) in Fig.~\ref{fig:Janossy});
we introduce a parameter ${z}$ so that $J_{k,p}(n_1,\ldots,n_k;{I})$ is given by
\begin{align}
J_{k,p}(n_1,\ldots,n_k;{I})&=\frac{1}{p!}\left(-\partial_{{z}}\right)^p
\det(\mathbb{I}-{z}\mathbf{K}_{I})\cdot \det\left[
\langle n_i|\mathbf{K}_{I}(\mathbb{I}-{z}\mathbf{K}_{I})^{-1}|n_j\rangle
\right]_{i,j=1}^k\bigg|_{{z}=1} ,
\label{Janossy_prob_p}
\end{align}
as the derivation of Eq.~(\ref{PkDet}) carries over to this case.
The  product of two determinants in Eq.{} (\ref{Janossy_prob_p}) can as well be written as
\begin{align}
(-1)^k\det
\left|
\begin{array}{cc}
-{\boldsymbol\kappa} & -\sqrt{{z}}\bs{k} \\
-\sqrt{{z}}\bs{k}^t & \mathbb{I}-{z}\mathbf{K}_I
\end{array}
\right|
=
\det{\boldsymbol\kappa}\cdot
\det\bigl(\mathbb{I}-
{z}\tilde{\mathbf{K}}_I
\bigr).
\label{Janossy_prob_p1}
\end{align}
Either by definition or from Eqs.~(\ref{Janossy_prob_p}) and (\ref{Janossy_prob_p1}),
the J\'{a}nossy density $J_{k,p}(n_1,\ldots,n_k;{I})$ reduces to:
(i) for $k=0$, the gap probability $E_0({I})=\det(\mathbb{I}-\mathbf{K}_{I})$
(or its generalization 
$E_p({I})=1/p!\,\left(-\partial_{{z}}\right)^p \left. \det(\mathbb{I}-{z}\mathbf{K}_{I})\right|_{{z}=1}$)
of finding no (or exactly $p$) particles in ${I}$, and 
(ii) formally for ${I}=\emptyset$ and $p=0$, to the $k$-point correlation function (\ref{Rk}).
Note also that the factor $\det{\boldsymbol\kappa}=\rho_k(n_1,\ldots,n_k)$ in
Eqs.~(\ref{Janossy_prob2}) and (\ref{Janossy_prob_p1})
is canceled when we consider the {\em conditional} probability $\tilde{J}_{k,p}(n_1,\ldots,n_k;{I})$
that there are exactly $p$ particles in a subset
${I}$ under the condition that $k$ particles are already at each of the $k$ designated loci, 
\begin{align}
\tilde{J}_{k,p}(n_1,\ldots,n_k;{I})=
\frac{1}{p!}\left.\left(-\partial_{{z}}\right)^p\det\bigl(\mathbb{I}-{z}\tilde{\mathbf{K}}_I\bigr)\right|_{{z}=1}.
\label{tildeJkp}
\end{align}

All the above formulas carry over to a continuous DPP on $\mathbb{R}$
and for a set of intervals $I\subset\mathbb{R}$.
Trivial modifications are needed to regard $\mathbf{K}$ and $\mathbf{K}_I$ as integral operators
\begin{align}
(\mathbf{K}\cdot f)(x)=\int dy \,K(x,y)f(y)
~~,~~
(\mathbf{K}_I \cdot f)(x)=\int_I dy \,K(x,y)f(y)
\end{align}
acting on the Hilbert space of square-integrable functions $L^2(\mathbb{R})$ and $L^2(I)$,
and to reinterpret joint probabilities $\rho_k, \tilde{\rho}_k, J_{k,p}, \tilde{J}_{k,p}$ 
as joint probability distributions.
For the  case of continuous DPPs
the expression $\<x|\mathbf{A}|y\>=A(x,y)$ (denoted as $\mathbf{A}\doteq A(x,y)$ in TW) means that
the integral operator $\mathbf{A}$ has a kernel equal to $A(x,y)$.
Namely, the J\'{a}nossy density $J_{k}(x_1,\ldots, x_k;{I})$
is defined as the probability density of finding exactly $k$ particles in ${I}$ 
and one at each of the $k$ infinitesimal intervals $(x_i,x_i+dx_i)\subset {I}$,
and is given by the Fredholm determinant $\Det(\mathbb{I}-\mathbf{K}_{I})$
times the (ordinary) 
determinant of the resolvent kernel of $\mathbf{K}_{I}$:
\begin{align}
J_{k}(x_1,\ldots, x_k;{I})
&=\Det(\mathbb{I}-\mathbf{K}_{I})\cdot \det\left[
\< x_i| \mathbf{K}_{I}(\mathbb{I}-\mathbf{K}_{I})^{-1} | x_j\>
\right]_{i,j=1}^k
\nonumber\\
&=
\det{\boldsymbol\kappa}\cdot
\Det\bigl(\mathbb{I}-\tilde{\mathbf{K}}_I\bigr)
~~,~~
\tilde{\mathbf{K}}=\mathbf{K}-\bs{k}^t {\boldsymbol\kappa}^{-1} \bs{k}.
\label{Janossy_density_cont}
\end{align}
Likewise its generalization $J_{k,p}(x_1,\ldots, x_k;{I})$ is given by
\begin{align}
J_{k,p}(x_1,\ldots, x_k;{I})=\det{\boldsymbol\kappa}\cdot\frac{1}{p!}\left.\left(-\partial_{{z}}\right)^p
\Det\bigl(\mathbb{I}-{z}\tilde{\mathbf{K}}_I\bigr)\right|_{{z}=1}.
\end{align}
Finally we note that the joint probability distribution of $k$ leftmost or rightmost particles is derived
from the J\'{a}nossy density for a semi-finite interval $I=(s_k, \infty)$ or $(-\infty, s_k)$ as
\begin{align}
P_{1\cdots k}(s_1,\ldots,s_{k})=
\left\{
\begin{array}{ll}
~~\partial_{s_k} J_{k-1}(s_1,\ldots, s_{k-1};(s_k,\infty)) & (s_1\geq \cdots \geq s_k)\\
-\partial_{s_k} J_{k-1}(s_1,\ldots, s_{k-1};(-\infty,s_k)) & (s_1\leq \cdots \leq s_k)
\end{array}
\right. .
\end{align}

\section{Applicability of the Tracy--Widom method}
\subsection{Inheritance of the Tracy--Widom criteria}
Consider a kernel of an integral operator $\mathbf{K}$ of the Christoffel--Darboux form
\begin{align}
K(x,y)=\frac{\varphi(x)\psi(y)-\psi(x)\varphi(y)}{x-y} , 
\label{CD}
\end{align}
with its component functions satisfying a pair of linear differential equations
\begin{align}
&m(x)\frac{d}{dx}
\left[
\begin{array}{c}
\varphi(x)\\
\psi(x)
\end{array}
\right]
=
\left[
\begin{array}{rr}
A(x) & B(x)\\
-C(x) & -A(x)
\end{array}
\right]
\left[
\begin{array}{c}
\varphi(x)\\
\psi(x)
\end{array}
\right]
\label{integrable}\\	
&~\mbox{with some polynomials}\ m,  A,  B, C.
\nonumber
\end{align}
The tracelessness of the $2\times 2$ matrix on the right-hand side of Eq.~(\ref{integrable})  is essential.
As a unifying approach to their preceding works on the sine \cite{Tracy:1993}, Airy \cite{Tracy:1994a} 
and Bessel kernels \cite{Tracy:1994b},
Tracy and Widom have shown in TW
that the Fredholm determinant $\mathrm{Det}(\mathbb{I}-\mathbf{K}_{I})$ of an operator $\mathbf{K}$ 
satisfying the  criteria (\ref{CD}), (\ref{integrable}) is always determined through a closed system of PDEs 
in the boundary points $\{a_i\}\in \partial I$.
This involves the boundary values of the functions
$Q_j(x)=((\mathbb{I}-\tilde{\mathbf{K}}_I)^{-1}\cdot x^j {\varphi})(x)$ and
$P_j(x)=((\mathbb{I}-\tilde{\mathbf{K}}_I)^{-1}\cdot x^j {\psi})(x)$,
and the inner products of $Q_j$ and $P_j$ with $\varphi$ and $\psi$ such as 
$u_j=\int_I dx \,\varphi(x) Q_j(x)$.
A large part of the TW system (Eqs.~(1.7a)--(1.9) and (2.12)--(2.18) of TW) is universal and the rest
(Eqs.~(2.25), (2.26) of TW) parametrically depends on the coefficients of the polynomials 
$m(x)=\sum_j \mu_j x^j$, $A(x)=\sum_j \alpha_j x^j$, etc.\\
\indent
Now we present a theorem:\\

\noindent
{\bf Theorem 1.}~
{\it If the kernel of $\mathbf{K}$ satisfies the TW criteria  (\ref{CD}), (\ref{integrable}),
so does the transformed kernel of $\tilde{\mathbf{K}}$.}\\

\noindent
{\it Proof.}~~
Since the kernel of $\tilde{\mathbf{K}}^{(k)}$
for the J\'{a}nossy density $J_{k}(x_1,\ldots,x_k;{I})$ is obtained from the kernel of $\tilde{\mathbf{K}}^{(k-1)}$ for
the J\'{a}nossy density  $J_{k-1}(x_1,\ldots,x_{k-1};{I})$ by adding an extra locus of particle $x_k=t$,
\begin{align}
\tilde{K}^{(k)}(x,y)=
\tilde{K}^{(k-1)}(x,y)-
\tilde{K}^{(k-1)}(x,t)
\tilde{K}^{(k-1)}(t,t)^{-1}
\tilde{K}^{(k-1)}(t,y),
\label{procedure}
\end{align}
by induction it is sufficient to prove Theorem 1 for $k=1$.
Then the transformed kernel is
\begin{align}
\tilde{K}(x,y)=K(x,y)-K(x,t)K(t,t)^{-1}K(t,y).
\label{tildeKx1}
\end{align}
Here we assumed that the density of particles $\rho_1(t)=K(t,t)=\varphi'(t)\psi(t)-\psi'(t)\varphi(t)$ 
at the designated locus $t$ is nonzero
(otherwise the J\'{a}nossy density would vanish by definition).
The transformed kernel  (\ref{tildeKx1}) is again of the Christoffel--Darboux form
\begin{align}
\tilde{K}(x,y)=
\frac{\tilde{\varphi}(x)\tilde{\psi}(y)-\tilde{\psi}(x)\tilde{\varphi}(y)}{x-y}~,
\end{align}
where 
\begin{align}
\begin{array}{l}
{\dps \tilde{\varphi}(x) =
{\varphi}(x)-\frac{K(x,t)}{K(t,t)}{\varphi}(t)=
 \varphi(x) - \frac{b(a \varphi(x) - b \psi(x))}{x-t}}\\
{\dps \tilde{\psi}(x)= 
{\psi}(x)-\frac{K(x,t)}{K(t,t)}{\psi}(t)=
\psi(x) - \frac{a(a \varphi(x) - b \psi(x))}{x-t}}
\end{array}
~~\mbox{with}~~
\begin{array}{l}
{\dps a:=\frac{\psi(t)}{\sqrt{\rho_1(t)}}}
\\
{\dps b:=\frac{\varphi(t)}{\sqrt{\rho_1(t)}}}
\end{array} .
\label{tildephipsi}
\end{align}
Using Eq.~(\ref{integrable}), they are shown to satisfy a set of linear differential equations
\begin{align}
&m(x)\frac{d}{dx}
\left[
\begin{array}{c}
\tilde\varphi(x)\\
\tilde\psi(x)
\end{array}
\right]
=
\left[
\begin{array}{rr}
\tilde{A}(x) & \tilde{B}(x)\\
-\tilde{C}(x) & -\tilde{A}(x)
\end{array}
\right]
\left[
\begin{array}{c}
\tilde\varphi(x)\\
\tilde\psi(x)
\end{array}
\right]
\end{align}
with
\begin{align}
\tilde{A}(x)&=
{A}(x)
+\frac{a^2 {B}(x)-b^2 {C}(x)}{x-t}
-\frac{a b \left(2 a b {A}(x)+a^2 {B}(x)+b^2 {C}(x)-m(x)\right)}{(x-t)^2},
\nonumber\\
\tilde{B}(x)&=
{B}(x)
-\frac{2b (b {A}(x)+a {B}(x))}{x-t}
+\frac{b^2 \left(2 a b {A}(x)+a^2 {B}(x)+b^2 {C}(x)-m(x)\right)}{(x-t)^2},
\label{ABC}
\\
\tilde{C}(x)&=
{C}(x)
+\frac{2 a (a {A}(x)+b{C}(x))}{x-t}
+\frac{a^2 \left(2 a b {A}(x)+a^2 {B}(x)+b^2 {C}(x)-m(x)\right)}{(x-t)^2}.
\nonumber
\end{align}
Since the coefficient functions $m$, ${A}, {B}$ and ${C}$ are polynomials in $x$,
so are the new coefficient functions after redefinition
$(x-t)^2 m(x)\mapsto m(x)$, 
$(x-t)^2\tilde{{A}}(x)\mapsto \tilde{{A}}(x)$, etc. \qed

\subsection{Conditioning particles' loci as gauge transformation}
Below we unravel the origin of inheritance of the TW criteria (\ref{CD}) and (\ref{integrable})
from $\mathbf{K}$ to  $\tilde{\mathbf{K}}$.
\begin{enumerate}
\item
The Christoffel--Darboux form (\ref{CD}):
suppose that $\mathbf{K}$ is composed of polynomials orthogonal with respect to a weight $w(x)$
or their asymptotic limits.
Then $\tilde{\mathbf{K}}$ is composed of polynomials orthogonal with respect to a weight
$\tilde{w}(x)= (x-t)^2 w(x)$ or their asymptotic limits, 
with the factor $(x-t)^2$ originating from the Vandermonde determinant squared.
\item
The tracelessness of the $2\times 2$ matrix in Eq.~(\ref{integrable}):
Eq.~(\ref{integrable})  specifies a covariantly constant section $\Psi(x)$ of
an $\mathbb{R}^2$-bundle over $\mathbb{R}$ with an 
$\mathfrak{sl}(2,\mathbb{R})$ connection $\mathcal{A}(x)$,
\begin{align}
&
(\partial_x +\mathcal{A}(x))\Psi(x)=0
~~,~~
\Psi(x)=
\left[
\begin{array}{l}
{\varphi}(x)\\
{\psi}(x)
\end{array}
\right],
\nonumber\\
&
\mathcal{A}(x)
=
-\frac{1}{m(x)}
\left[
\begin{array}{rr}
{A}(x) & {B}(x)\\
-{C}(x) & -{A}(x)
\end{array}
\right]
~~\mbox{satisfying}~~
\mathrm{tr}\,\mathcal{A}(x)=0,
\end{align}
and
Eq.~(\ref{tildephipsi}) is an $\mathrm{SL}(2,\mathbb{R})$ gauge transformation,
\begin{align}
&
\tilde{\Psi}(x)=\mathcal{U}(x)\Psi(x)
~~,~~
\tilde{\Psi}(x)=
\left[
\begin{array}{l}
\tilde{\varphi}(x)\\
\tilde{\psi}(x)
\end{array}
\right],
\nonumber\\
&
\mathcal{U}(x)=
\left[
\begin{array}{rr}
{\dps  1-\frac{a b}{x-t}} & {\dps \frac{b^2}{x-t}} \\
{\dps  -\frac{a^2}{x-t}} & {\dps 1+\frac{a b}{x-t}}
\end{array}
\right]
~~\mbox{\em satisfying}~~
\det \mathcal{U}(x)=1.
\end{align}
Then the gauge-transformed section $\tilde{\Psi}(x)$ must be covariantly constant
\begin{align}
&
(\partial_x +\tilde{\mathcal{A}}(x))\tilde{\Psi}(x)=0
~~,~~
\tilde{\mathcal{A}}(x)=\mathcal{U}(x) \mathcal{A}(x) \mathcal{U}(x)^{-1}-
\partial_x \mathcal{U}(x) \cdot \mathcal{U}(x)^{-1}
\label{gauge}
\end{align}
for the gauge-transformed $\mathfrak{sl}(2,\mathbb{R})$ connection $\tilde{A}(x)$ {\em that remains traceless},
$\mathrm{tr}\,\tilde{\mathcal{A}}(x)=0$.
Repetition of gauge transformations of the form
\begin{align}
\mathcal{U}(x)=
\left[
\begin{array}{rr}
{\dps  1-\frac{a_k b_k}{x-x_k}} & {\dps \frac{b_k^2}{x-x_k}} \\
{\dps  -\frac{a_k^2}{x-x_k}} & {\dps 1+\frac{a_k b_k}{x-x_k}}
\end{array}
\right]
\cdots
\left[
\begin{array}{rr}
{\dps  1-\frac{a_1 b_1}{x-x_1}} & {\dps \frac{b_1^2}{x-x_1}} \\
{\dps  -\frac{a_1^2}{x-x_1}} & {\dps 1+\frac{a_1 b_1}{x-x_1}}
\end{array}
\right]
\end{align}
on ${\Psi}(x)$ yields the $k$th-order J\'{a}nossy density $J_k(x_1,\ldots,x_k;I)$.
Although 
the gauge transformation
$\mathcal{U}(x)$ has poles at
$x=x_1,\ldots, x_k$, 
the transformed section $\tilde{\Psi}(x)$ is regular and vanishes there.
\item
Meromorphy of ${\mathcal{A}}(x)$ inherits down to $\tilde{\mathcal{A}}(x)$ by Eq.~(\ref{gauge})
(which is equivalent to Eq.{} (\ref{ABC})),
as $\mathcal{U}(x)$ is meromorphic.
\end{enumerate}

Accordingly, the TW method is applicable to the evaluation of J\'{a}nossy densities
of any continuous DPP if it is applicable to the evaluation of its gap probability,
and $J_{k,p}(x_1,\ldots,x_k;I)$ is expressible in terms of a
solution to a system of partial differential equations (PDEs)
(containing $x_1,\ldots,x_k$ parametrically) in the endpoints $\{a_i\}$ of $I$.\\
\indent
A few comments are in order:
\begin{itemize}
\item
By construction (\ref{tildeKx1}), the transformed kernel $\tilde{K}(x,y)$ vanishes when one of the arguments is equal to $t$,
\begin{align}
\tilde{K}(x,t)=\tilde{K}(t,y)=0.
\label{tildeKxt}
\end{align}
This leads to,  for ${}^\forall  f\in L^2(I)$,
\begin{align}
(\tilde{\mathbf{K}}_I\cdot f)(t)=0\ \ \ 
\mbox{and thus}\ \ \ 
((\mathbb{I}-\tilde{\mathbf{K}}_I)^{-1}\cdot f)(t)=f(t).
\label{tildeKft0}
\end{align}
\item
As mentioned above,
the functions $\tilde{\varphi}(x)$ and $\tilde{\psi}(x)$ 
also vanish at
$x=t$.
It leads to
\begin{align}
q_j(t):=((\mathbb{I}-\tilde{\mathbf{K}}_I)^{-1}\cdot x^j \tilde{\varphi})(t)=
t^j\tilde{\varphi}(t)=0,
\nonumber\\
p_j(t):=((\mathbb{I}-\tilde{\mathbf{K}}_I)^{-1}\cdot x^j\tilde{\psi})(t)=
t^j\tilde{\psi}(t)=0
\label{qpt}
\end{align}
for $j\in \mathbb{N}$.
These could serve as part of the boundary conditions for the TW system, but
we later use them only for a consistency check of the solution $q_j(s)$ and $p_j(s)$
derived from a different boundary condition imposed either at $s\gg1 $ or $s\ll 1$.
\item
For the sine kernel ${\dps K(x,y)=\frac{\sin(x-y)}{\pi(x-y)}
=\frac{\sqrt{xy}}{2}\frac{J_{1/2}(x)J_{-1/2}(y)-J_{-1/2}(x)J_{1/2}(y)}{x-y}
}$
governing the spectral bulk of unitary ensembles,
Forrester and Odlyzko \cite{Forrester:1996} previously considered 
the Fredholm determinant of
its transformed kernel
\cite{Nagao:1993}
(which they denoted as $K_1$ instead of our $\tilde{K}$)
with $k=1$ and $t$ set to 0 without loss of generality,
\begin{align}
{K}_1(x,y)
=\frac{\sqrt{xy}}{2}
\frac{J_{3/2}(x)J_{1/2}(y)-J_{1/2}(x)J_{3/2}(y)}{x-y}
=\frac{1}{\pi}
\(\frac{\sin(x-y)}{x-y}-\frac{\sin x}{x}\frac{\sin y}{y}\).
\label{K1}
\end{align}
They did apply the TW method to $K_1(x,y)$ and expressed the Fredholm determinant on a symmetric interval
$\mathrm{Det}(\mathbb{I}-\mathbf{K}_1|_{(-s,s)})$ in terms of a solution to the TW system of 
ordinary differential equations (ODEs).
However, in order to invoke the TW method
they paid attention to an apparent fact that $K_1$ is related to $K$ by a unit shift of the indices of the Bessel functions,
rather than by an $\mathrm{SL}(2,\mathbb{R})$ gauge transformation
$
\mathcal{U}(x)=
\left[
{
~~1~~~\;0
\atop 
-x^{-1} \ 1
}
\right]
$
that retains the tracelessness of 
$\mathcal{A}(x)=
\left[
{
0\; -1
\atop 
1~\ \,0
}
\right].
$
Nor did they explicitly write $K_1$ in a form of the right-hand side of Eq.~(\ref{K1}), which would have meant
$K(x,y)-K(x,0)K(0,0)^{-1}K(0,y)$.
Our formulation is 
a systematic generalization of the spirit of their work to arbitrary kernels of the TW type,
to any interval $I$, and to any number ($k\geq 2$) of conditioned particles.

\end{itemize}

\section{Applications to random matrix theory}
In this section we consider a DPP of 
eigenvalues $\{x_i\}$ of an $N\times N$ unitary-invariant random matrix ensemble with measure
\begin{align}
\prod_{i=1}^N dx_i\, w(x_i) \cdot \prod_{i>j}^N (x_i-x_j)^2.
\end{align} 
There the functions $\varphi(x)$ and $\psi(x)$ are (asymptotic forms of)
the $N$th and $(N-1)$th of the
polynomials orthogonal with respect to the weight $w(x)$.
In this case, the conditional probability distributions
$\tilde{\rho}_p(x_1,\ldots,x_p; t_1,\ldots,t_k)$ (\ref{tilderhop}) and
$\tilde{J}_{k,p}(t_1,\ldots,t_k; I)$ (\ref{tildeJkp})
described by the transformed kernel $\tilde{\mathbf{K}}$
are nothing other than  {\em unconditional} probability distributions
of eigenvalues of a random matrix ensemble with weight function
$\tilde{w}(x)=w(x)\prod_{j=1}^k (x-t_j)^2$, and 
$\tilde{\varphi}(x)$ and $\tilde{\psi}(x)$ are (asymptotic forms of)
polynomials orthogonal with respect to $\tilde{w}(x)$.\footnote{
This fact was previously used
to compute the $p$-point correlation functions of the
``massive" Bessel and sine kernels
corresponding to the microscopic scaling limit of unitary-invariant random matrix 
ensembles with weights
$w(x)=\e^{-x}x^\nu \prod_{j=1}^k (x+m^2_j)\Theta(x)$ \cite{Damgaard:1998a} and 
$w(x)=\e^{-x^2}\prod_{j=1}^k (x^2+m_j^2)$ \cite{Damgaard:1998b} (see, e.g.,~Eq.~(33) of Ref.~\cite{Damgaard:1998a}),
as effective models of 4- and 3D QCD with $k$ (pairs of) dynamical quarks of masses $\{m_j\}$.}

If the values of the resolvent kernel
$R(x,y)=\< x|\mathbf{K}_{I}(\mathbb{I}-\mathbf{K}_{I})^{-1}| y\>$ for arbitrary $x, y\in I$ 
(not just its boundary values $R(a_i, a_j)$ at $a_i,a_j \in  \partial I$, as derived in TW)
were analytically available for various kernels appearing in the RMT, the
J\'{a}nossy densities would  readily be computed from the first line of Eq.~(\ref{Janossy_density_cont}).
Since this path is infeasible 
(despite the fact that numerical evaluation of $\< x|\mathbf{K}_{I}(\mathbb{I}-\mathbf{K}_{I})^{-1}| y\>$
or $\mathrm{Det}(\mathbb{I}-\tilde{\mathbf{K}}_{I})$
by the quadrature approximation is always possible \cite{Nishigaki:2016}),
we choose the second line of Eq.~(\ref{Janossy_density_cont}) as an alternate route.
As illustrative examples, we compute J\'{a}nossy densities for the Airy and Bessel kernels
by applying the TW method to their transformed kernels.

\subsection{J\'{a}nossy density for the Airy kernel}
The Airy kernel governs local fluctuation and correlation of scaled eigenvalues of
random Hermitian $N\times N$ matrices
at the soft edge where the {\em global} density descends to zero as $ (-x)^{1/2}$.
It consists of
\begin{align}
\varphi(x)=\mathrm{Ai}(x)
,~~
\psi(x)=\mathrm{Ai}'(x)
\end{align}
from which it follows that
\begin{align}
m(x)=1,~~
A(x)=0,~~
B(x)=1,~~
C(x)=-x.
\end{align}
As an example we concentrate on the simplest of J\'{a}nossy densities, $J_{1}(t; I)$ with $I=(s,\infty)$
and already set ${z}$ to unity.
Note that ${P}_{12}(t,s)=\Theta(t-s)\partial_s J_{1}(t; (s,\infty))$ represents the joint distribution 
of the first and second largest eigenvalues $(t,s)$ of unitary ensembles,
previously derived in Ref.~\cite{Witte:2013} via Ref.~\cite{Forrester:2007}  
using a much more elaborate method than this work.

The coefficient functions (\ref{ABC}), whose degrees are increased by two after the redefinition
$(x-t)^2 m(x)\mapsto m(x)$, $(x-t)^2\tilde{{A}}(x)\mapsto \tilde{{A}}(x)$, etc., read
\begin{align}
m(x)&=
(x-t)^2
&
\nonumber\\
\tilde{A}(x)&=
-a b(a^2-1) -a^2 t
+ \left(a^2+a b^3-b^2 t\right)x+b^2 x^2
&:=\sum\nolimits_{j=0}^2 \alpha_j x^j
\nonumber\\
\tilde{B}(x)&=
b^2(a^2 -1)
+2 a b t+t^2- \left(2 a b+b^4+2 t\right)x+x^2
&:=\sum\nolimits_{j=0}^2 \beta_j x^j
\label{ABC_Airy}
\\
\tilde{C}(x)&=
a^2 (a^2-1)- (a b-t)^2 x-2 (a b-t)x^2 -x^3
&:=\sum\nolimits_{j=0}^3 \gamma_j x^j
\nonumber
\end{align}
with
\begin{align}
a=\frac{\mathrm{Ai}'(t)}{\sqrt{\rho_1(t)}}~,~
b=\frac{\mathrm{Ai}(t)}{\sqrt{\rho_1(t)}}~,~
\rho_1(t)=\mathrm{Ai}'(t)^2-t\, \text{Ai}(t)^2.
\end{align}
Equation (\ref{ABC_Airy}) could be slightly simplified by using the relation $a^2-b^2 t=1$ but we refrain.
By taking the right endpoint $a_2$ of $I$ to $+\infty$,
all terms in the TW system that contain $a_2$ vanish because of the exponential decay of the Airy kernel.
Then the quantities involved,
\begin{align}
&
R(s)=\langle s| \tilde{\mathbf{K}}_{I}(\mathbb{I}-\tilde{\mathbf{K}}_{I})^{-1}|s\rangle\ \ 
(\mbox{abbreviated notation of }R(s,s)),
\\
&
q_k(s)=((\mathbb{I}-\tilde{\mathbf{K}}_{I})^{-1} \cdot x^k\tilde\varphi)(s),\ \  
p_k(s)=((\mathbb{I}-\tilde{\mathbf{K}}_{I})^{-1} \cdot x^k\tilde\psi)(s),
\nonumber\\
&
u_k(s)=\int_I dx\, \tilde\varphi(x)\,x^k ((\mathbb{I}-\tilde{\mathbf{K}}_{I})^{-1} \cdot \tilde\varphi)(x),\ \ 
v_k(s)=\int_I dx\, \tilde\psi(x)\,x^k ((\mathbb{I}-\tilde{\mathbf{K}}_{I})^{-1} \cdot \tilde\varphi)(x),
\nonumber\\
&
\tilde{v}_k(s)=\int_I dx\, \tilde\varphi(x)\,x^k ((\mathbb{I}-\tilde{\mathbf{K}}_{I})^{-1}\cdot  \tilde\psi)(x),\ \ 
w_k(s)=\int_I dx\, \tilde\psi(x)\,x^k ((\mathbb{I}-\tilde{\mathbf{K}}_{I})^{-1}\cdot  \tilde\psi)(x),
\nonumber
\end{align}
are all treated as functions of the left endpoint $a_1=s$ alone, and their parametric dependence on $t$ is implicit.
The  system of ODEs (Eqs.~(2.25)--(2.26), (2.15)--(2.18), (2.12)--(2.14) of TW)  takes the form 
($'=\partial_s$, and the arguments $(s)$ are suppressed):
\begin{align}
(s-t)^2{}\,q_0'{}
&=
\sum_{j=0}^2\(\alpha_j+\sum_{k=0}^1 \alpha_{j+k+1}v_k{}+\sum_{k=0}^2 \gamma_{j+k+1}u_k{}\)q_j{}
-v_0{} q_0{}
\nonumber\\
&+
\sum_{j=0}^2\(\beta_j+\sum_{k=0}^1 \alpha_{j+k+1}u_k{}+\sum_{k=0}^1 \beta_{j+k+1}v_k{}\)p_j{}
+u_0{} p_0{}~,
\nonumber\\
(s-t)^2{}\,p_0'{}
&=
\sum_{j=0}^3\(-\gamma_j+\sum_{k=0}^1 \alpha_{j+k+1}w_k{}+\sum_{k=0}^2 \gamma_{j+k+1}\tilde{v}_k{}\)q_j{}
-w_0{} q_0{}
\nonumber\\
&+
\sum_{j=0}^2\(-\alpha_j+\sum_{k=0}^1 \alpha_{j+k+1}\tilde{v}_k{}+\sum_{k=0}^1 \beta_{j+k+1}w_k{}\)p_j{}
+\tilde{v}_0{} p_0{}~,
\label{longeqsA}
\end{align}
\begin{align}
u_0'&=-q_0q_0~,~
u_1'=-q_0q_1~,~
u_2'=-q_0q_2~,\nonumber\\
v_0'&=-q_0p_0~,~
v_1'=-q_0p_1~,~
v_2'=-q_0p_2~,\nonumber\\
w_0'&=-p_0p_0~,~
w_1'=-p_0p_1~,\nonumber\\
q_1{}&=
s\,q_0{}-v_0{}q_0{}+u_0{}p_0{}~,
\nonumber\\
q_2{}
&=
s^2q_0{}-v_0{}q_1{}-v_1{}q_0{}+u_0{}p_1{}+u_1{}p_0{}~,
\nonumber\\
q_3{}
&=
s^3q_0{}-v_0{}q_2{}-v_1{}q_1{}-v_2{}q_0{}+u_0{}p_2{}+u_1{}p_1{}+u_2{}p_0{}~,
\nonumber\\
p_1{}
&=
s\,p_0{}-w_0{}q_0{}+\tilde{v}_0{}p_0{}~,
\nonumber\\
p_2{}
&=
s^2p_0{}-w_0{}q_1{}-w_1{}q_0{}+\tilde{v}_0{}p_1{}+\tilde{v}_1{}p_0{}~,
\nonumber\\
\tilde{v}_0&=v_0~,
\nonumber\\
\tilde{v}_1&=v_1-v_0\tilde{v}_0+u_0 w_0~,
\nonumber\\
\tilde{v}_2&=v_2-v_0\tilde{v}_1-v_1\tilde{v}_0+u_0 w_1+u_1 w_0~.
\nonumber
\end{align}
The exponential decay of the Airy kernel and thus the transformed kernel
($\log \tilde{K}(x,y)\simeq\log{K}(x,y)\simeq -\frac23 x^{3/2}$ for $x\gg1$ and $y, t$ fixed) 
leads to the boundary conditions for $s\gg 1$:
\begin{align}
q_0(s)&\simeq \tilde\varphi(s)~,~
p_0(s)\simeq \tilde\psi(s)~,
\nonumber
\\
u_k(s) &\simeq \int_s^\infty dx\,x^k \tilde\varphi(x)^2~,~
v_k(s) \simeq \int_s^\infty dx\,x^k \tilde\varphi(x)\tilde\psi(x)~,~
w_k(s) \simeq \int_s^\infty dx\,x^k \tilde\psi(x)^2~.
\label{ukvkwks}
\end{align}
The diagonal resolvent (Eq.~(1.7b) of TW) and  the Fredholm determinant
of the transformed kernel $\tilde{\mathbf{K}}_{I}$ are expressed in terms of  the solution to the ODEs (\ref{longeqsA}):
\begin{align}
&R(s)=\partial_s \log \mathrm{Det}(\mathbb{I}-\tilde{\mathbf{K}}_{(s,\infty)})
=p_0(s) q_0'(s)-q_0(s) p_0'(s) .
\label{RsAiry}
\end{align}

\begin{figure}[t] 
\begin{center}
\hspace{1cm}
\includegraphics[bb=0 0 502 343,width=80mm]{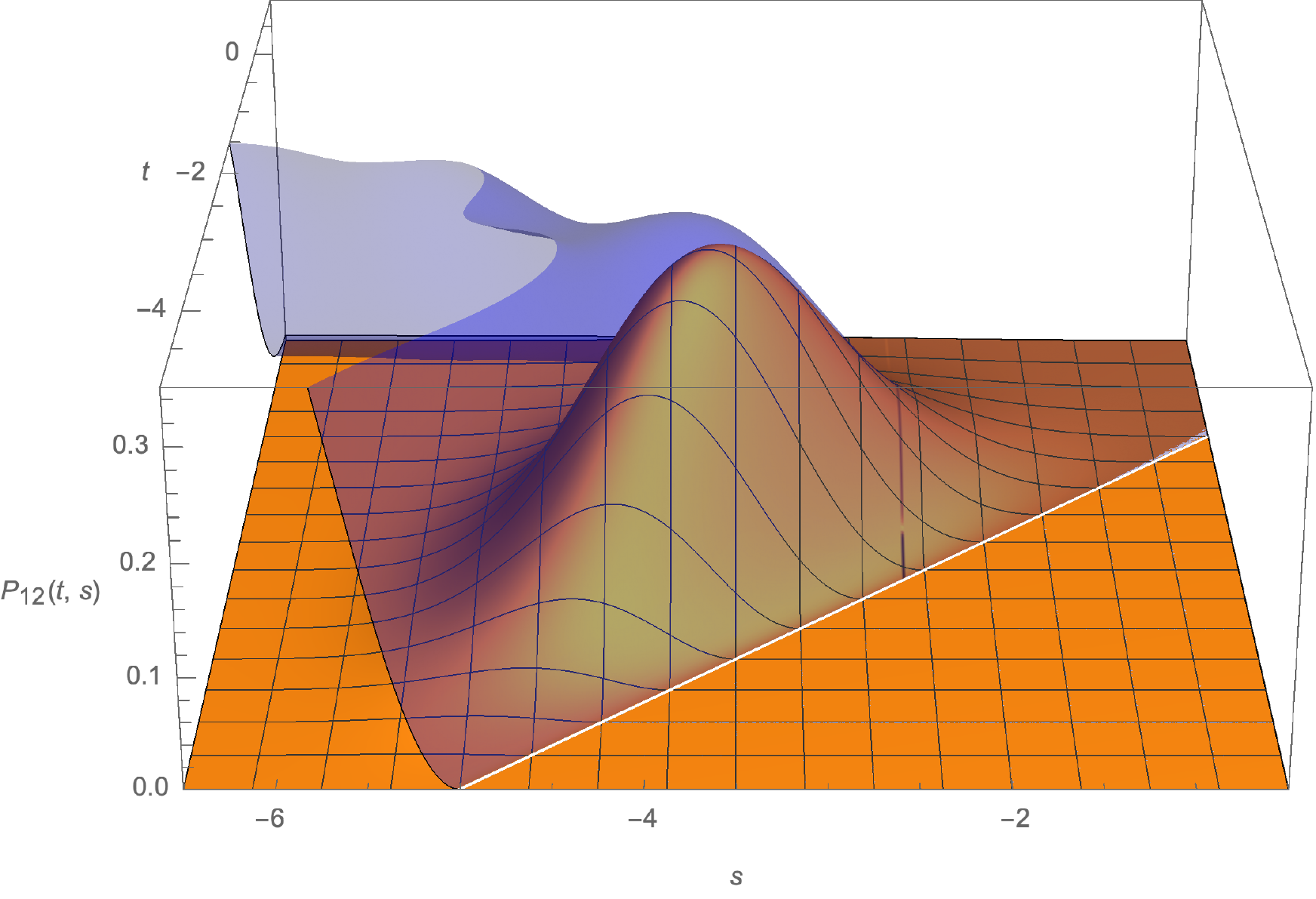}
\caption{\label{fig:P12AiryJD}
The joint distribution 
of the first and second largest eigenvalues ${P}_{12}(t,s)$ (orange) and
the two-point correlation function ${\rho}_2(t, s)$ for $t>s$ (transparent blue) of random Hermitian matrices.}
\end{center}
\end{figure}

For numerical evaluation of the solution, in practice we impose the boundary condition
$q_0(\Lambda)= \varphi(\Lambda)$, etc., at a sufficiently large positive $\Lambda$ $(\sim 10)$.
Since $q_0(s)$ and $p_0(s)$ are  regular at $s=t$ (they are actually zero by Eq.~(\ref{qpt})),
apparent ``double poles" at $s=t$ in the first two nonuniversal equations of Eq.~(\ref{longeqsA}) 
are guaranteed to be canceled by the double zeroes on the right-hand side.
Nevertheless, this could potentially cause loss of numerical accuracy
when solving the TW system of ODEs from $s=\Lambda$ down to $s<t$, e.g.,~by the explicit Runge--Kutta method.
We have verified that
this apparent stiffness at $s=t$ can be circumvented by adding 
to $t$ a tiny imaginary part $\epsilon$ of the order of $O(10^{-10})$. With appropriately chosen values
of $\epsilon=\Im m(t)$, the real parts of $q_0(s)$ and $p_0(s)$ are stable upon varying $\epsilon$, and $q_0(\Re e(t))$
and $p_0(\Re e(t))$ vanish up to the accuracy of $O(\epsilon)$ as they should. In Fig.~\ref{fig:P12AiryJD} we display the joint
distribution of the largest eigenvalue $t$ and the second largest eigenvalue $s$
\begin{align}
{P}_{12}(t,s)=
\Theta(t-s)
\partial_s \(
\rho_1(t)
\mathrm{Det}(\mathbb{I}-\tilde{\mathbf{K}}_{(s,\infty)})\)
=\Theta(t-s)
\rho_1(t) R(s)\exp\(-\int_s^\infty ds'R(s')\)
\end{align}
obtained by this prescription.
The two-point correlation function 
$\rho_2(t,s)=\rho_1(t)\rho_1(s)-K(t,s)^2$, which is composed of peaks of
joint distributions ${P}_{k\ell}(t,s)$ of the $k$th and $\ell$th largest eigenvalues for $t>s$ ($k<\ell$),
is overlaid for comparison.
We have checked that the Fredholm determinant 
$\mathrm{Det}(\mathbb{I}-\tilde{\mathbf{K}}_{(s,\Lambda)})=\exp\(-\int_s^\Lambda ds'R(s')\)$
obtained by the TW system
is in perfect agreement with numerical values from
the Nystr\"{o}m-type quadrature approximation \cite{Bornemann:2010a,Bornemann:2010b}
\begin{align}
\mathrm{Det}(\mathbb{I}-\tilde{\mathbf{K}}_{(s,\Lambda)})
\simeq \det \left[ \delta_{ab} -\tilde{K}(x_a,x_b)\sqrt{w_a w_b} \right]_{a,b=1}^M~,
\label{Nystrom}
\end{align}
where $(x_1,\ldots,x_M; w_1,\ldots,w_M)$ is the Gauss--Legendre quadrature of the interval $(s,\Lambda)$.
Specifically, for a cutoff value $\Lambda=10$,
relative deviations between 
$\int_s^\Lambda ds'R(s')$ computed from the TW system (\ref{longeqsA}), (\ref{ukvkwks}), (\ref{RsAiry}) 
using Mathematica's {\tt NIntegrate} (for the preparation of boundary values) and {\tt NDSolve} (for solving coupled ODEs)
with $\epsilon=10^{-12}$ and at  quadruple {\tt WorkingPrecision},
and $-\log\mathrm{Det}(\mathbb{I}-\tilde{\mathbf{K}}_{(s,\Lambda)})$ 
computed by the Nystr\"{o}m-type approximation (\ref{Nystrom}) with quadrature order $M=200$,
are between $10^{-11}$ and $10^{-8}$ for a range of variables $-7\leq s,t\leq 5$
(see Table 1 for $t=-2$ and $s=-7,\ldots,5$).
The table of numerical data for  the Fredholm determinant 
$\mathrm{Det}(\mathbb{I}-\tilde{\mathbf{K}}_{(s,\infty)})$
is attached as online supplementary material.

\begin{table}[h]
\centering
\begin{tabular}{ccccccc}
\hline
{\small $s$} & {\small $-7$} & {\small $-6$} & {\small $-5$} & {\small $-4$} & {\small $-3$} & {\small $-2$}\\
{\small Rel. dev.} & 
{\small $\!\!4.20\times 10^{-11}\!\!$} & {\small $\!\!-4.41\times 10^{-12}\!\!$} & 
{\small $\!\!-7.72\times 10^{-11}\!\!$} & {\small $\!\!8.93\times 10^{-11}\!\!$} &
{\small $\!\!9.12\times 10^{-10}\!\!$} & {\small $\!\!-2.33\times 10^{-10}\!\!$} \\
\hline
{\small $-1$} & {\small 0} & {\small 1} & {\small 2} & {\small 3} & {\small 4} & {\small 5} \\
{\small $\!\!-7.33\times 10^{-9}\!\!$} & {\small $\!\!2.39\times 10^{-11}\!\!$} &
{\small $\!\!1.98\times 10^{-10}\!\!$} & {\small $\!\!6.65\times 10^{-10}\!\!$} &
{\small $\!\!2.94\times 10^{-9}\!\!$} & {\small $\!\!3.08\times 10^{-10}\!\!$} & 
{\small $\!\!2.55\times 10^{-9}\!\!$}\\
\hline
\end{tabular}
\caption{Relative deviations of $\log\mathrm{Det}(\mathbb{I}-\tilde{\mathbf{K}}_{(s,10)})$ for the Airy kernel
at $t=-2$ and various values of $s$
computed by the TW method ($\Im m(t)=10^{-12}$, 
{\tt WorkingPrecision $\mathtt{\to}$ 4 MachinePrecision}) 
versus the ones by the Nystr\"{o}m-type approximation ($M=200$).}
\end{table}

\subsection{J\'{a}nossy density for the Bessel kernel}
The Bessel kernel governs
local fluctuation and correlation of scaled eigenvalues of random 
positive-definite $N\times N$ Hermitian matrices $H$
at the hard edge where the weight function behaves as
$w(x)\simeq x^\nu \Theta(x)\ (\nu>-1).$
Equivalently it also governs local fluctuation of near-zero singular values of random complex
$N\times (N+\nu)$ matrices $W$ 
(i.e.,~the square root of the eigenvalues of Wishart matrices $H=W^\dagger W$)
by a redefinition of variables $x\mapsto x^2$.
It consists of
\begin{align}
\varphi(x)=J_\nu(\sqrt{x}),~~
\psi(x)=\frac{ \sqrt{x}}{4}\left(J_{\nu-1}(\sqrt{x})-J_{\nu+1}(\sqrt{x})\right),
\end{align}
from which it follows that
\begin{align}
m(x)=x,~~
A(x)=0,~~
B(x)=1,~~
C(x)=\frac14(x-\nu^2).
\end{align}
Again we concentrate on the simplest of J\'{a}nossy densities, $J_{1}(t; I)$ with $I=(0,s)$,
and already set ${z}$ to unity.
${P}_{12}(t,s)=-\Theta(t-s)\partial_s J_{1}(t; (s,\infty))$ represents the joint distribution 
of the first and second smallest eigenvalues $(t,s)$ of unitary ensembles,
previously derived in Ref.~\cite{Forrester:2007} using more elaborate methods.
The new coefficient functions in Eq.~(\ref{ABC}) are, after redefinition,
\begin{align}
{m}(x)&=
x(x-t)^2
&
\nonumber\\
\tilde{A}(x)&=
-a^2 (a b+t)+
\frac{b^2 \nu ^2}{4}  (a b-t)
+
\(
a^2+a b-\frac{b^2}{4}  (a b-t)+\frac{b^2 \nu ^2}{4}
\)x
-\frac{b^2 }{4}x^2
\nonumber\\
&
:=\sum\nolimits_{j=0}^2 \alpha_j x^j
\nonumber\\
\tilde{B}(x)&=
(a b+t)^2-\frac{b^4 \nu ^2}{4}
+
\(
-b^2+\frac{b^4}{4}-2 (a b+t)
\)x
+x^2
\nonumber\\
&:=\sum\nolimits_{j=0}^2 \beta_j x^j
\label{ABC_Bessel}
\\
\tilde{C}(x)&=
a^4-\frac{\nu ^2}{4} (a b-t)^2
+
\(
-a^2
+\frac{1}{4} (a b-t)^2
-\frac{\nu ^2}{2}  (a b-t)
\)x
+
\(\frac{1}{2} (a b-t)-\frac{\nu^2}{4}\)x^2+
\frac{x^3}{4}
\nonumber\\
&:=\sum\nolimits_{j=0}^3 \gamma_j x^j
\nonumber
\end{align}
with
\[
a=\frac{ \sqrt{t}\left(J_{\nu-1}(\sqrt{t})-J_{\nu+1}(\sqrt{t})\right)}{4\sqrt{\rho_1(t)}}
~,~
b=\frac{J_\nu(\sqrt{t})}{\sqrt{\rho_1(t)}}
~,~
\rho_1(t)=\frac14\left(J_\nu(\sqrt{t}){}^2-
J_{\nu-1}(\sqrt{t})   J_{\nu+1}(\sqrt{t})\right).
\]
The TW system of ODEs again takes the form (\ref{longeqsA}), with the first two nonuniversal equations replaced by
\begin{align}
s(s-t)^2{}\,q_0'{}
&=
\sum_{j=0}^2\(\alpha_j+\sum_{k=0}^1 \alpha_{j+k+1}v_k{}+\sum_{k=0}^2 \gamma_{j+k+1}u_k{}\)q_j{}
+2t\, v_0 q_0-2v_1q_0-v_0q_1~,
\nonumber\\
&+
\sum_{j=0}^2\(\beta_j+\sum_{k=0}^1 \alpha_{j+k+1}u_k{}+\sum_{k=0}^1 \beta_{j+k+1}v_k{}\)p_j{}
-2t\, u_0 p_0+2u_1p_0+u_0p_1~,
\label{longeqsB}\\
s(s-t)^2{}\,p_0'{}
&=
\sum_{j=0}^3\(-\gamma_j+\sum_{k=0}^1 \alpha_{j+k+1}w_k{}+\sum_{k=0}^2 \gamma_{j+k+1}\tilde{v}_k{}\)q_j{}
+2t\, w_0 q_0-2w_1q_0-w_0q_1~,
\nonumber\\
&+
\sum_{j=0}^2\(-\alpha_j+\sum_{k=0}^1 \alpha_{j+k+1}\tilde{v}_k{}+\sum_{k=0}^1 \beta_{j+k+1}w_k{}\)p_j{}
-2t\, \tilde{v}_0 p_0+2\tilde{v}_1p_0+\tilde{v}_0p_1~.
\nonumber
\end{align}
and the next eight universal equations sign-flipped:
\begin{align}
u_0'=q_0q_0,
u_1'=q_0q_1,
u_2'=q_0q_2,
v_0'=q_0p_0,
v_1'=q_0p_1,
v_2'=q_0p_2,
w_0'=p_0p_0,
w_1'=p_0p_1.
\label{signflip}
\end{align}
Note that by setting the left endpoint $a_1$ of $I$ to $0$, all terms containing $a_1$ either vanish or decouple.
Accordingly all quantities are treated as functions of the right endpoint $s$
alone, and their parametric dependence on $t$ is implicit.
Boundary conditions for $s\ll 1$ are:
\begin{align}
q_0(s)&\simeq \tilde\varphi(s)~,~
p_0(s)\simeq \tilde\psi(s)~,
\nonumber
\\
u_k(s) &\simeq \int_0^s dx\,x^k \tilde\varphi(x)^2~,~
v_k(s) \simeq \int_0^s dx\,x^k \tilde\varphi(x)\tilde\psi(x)~,~
w_k(s) \simeq \int_0^s dx\,x^k \tilde\psi(x)^2~.
\label{ukvkwksB}
\end{align}
The diagonal resolvent  and  the Fredholm determinant
of the transformed kernel $\tilde{\mathbf{K}}_{I}$ are expressed in terms of the solution to the ODEs (\ref{longeqsB}):
\begin{align}
R(s)=-\partial_s \log \mathrm{Det}(\mathbb{I}-\tilde{\mathbf{K}}_{(0,s)})=p_0(s) q_0'(s)-q_0(s) p_0'(s) .
\label{RsBessel}
\end{align}

\begin{figure}[b] 
\begin{center}
\hspace{1cm}
\includegraphics[bb=0 0 360 242,width=76mm]{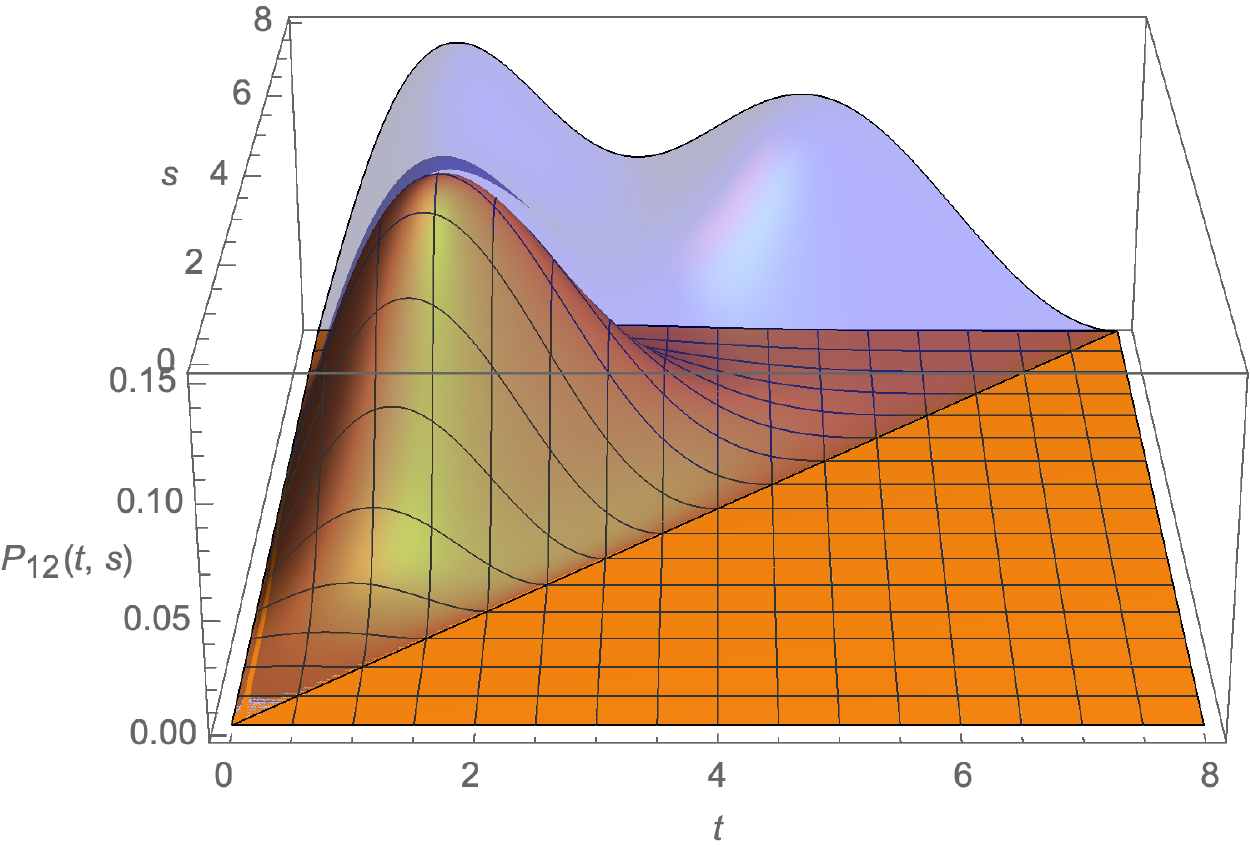}~~
\includegraphics[bb=0 0 360 247,width=76mm]{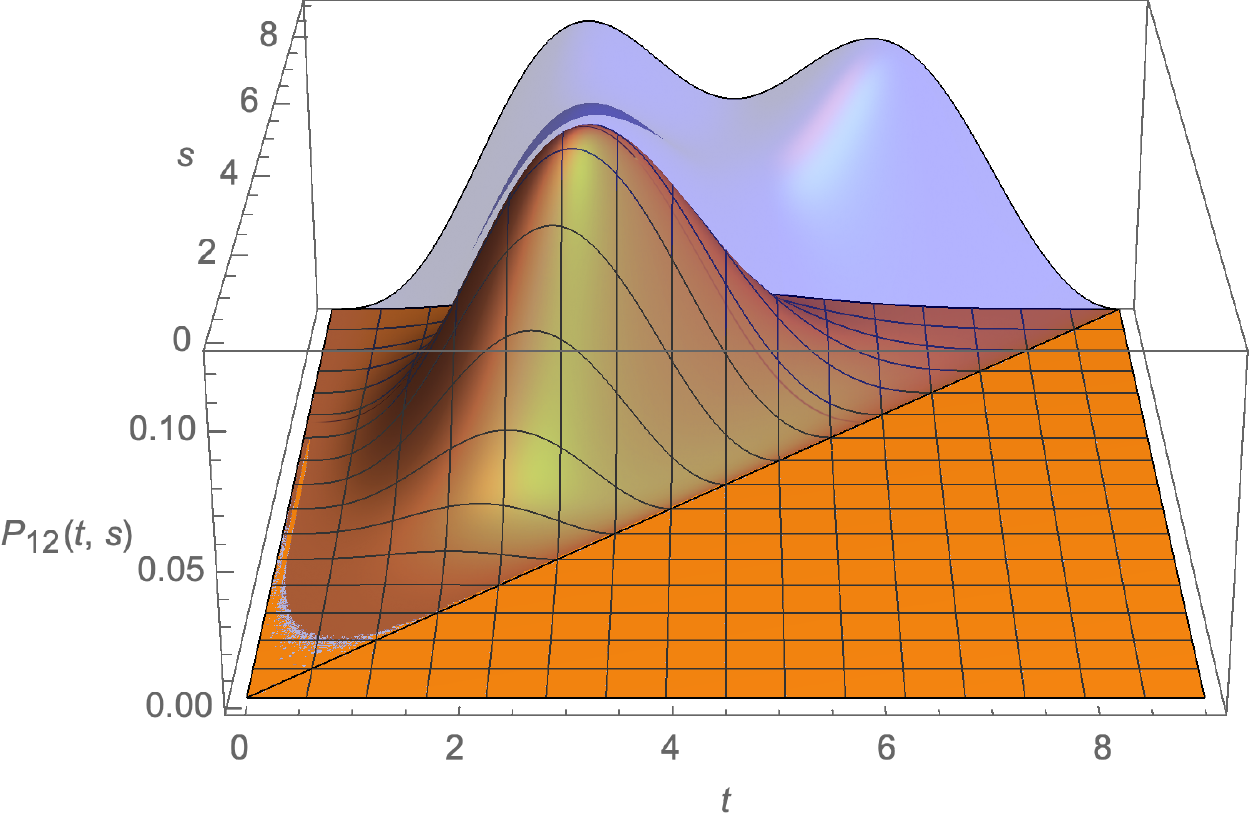}
\caption{\label{fig:P12BesselJD}
The joint distribution of the first and second largest singular values ${P}_{12}(t,s)$ (orange) and
the two-point correlation function ${\rho}_2(t, s)$ for $t>s$ (transparent blue)
of random complex matrices with $\nu=0$ (left) and $\nu=1$ (right).}
\end{center}
\end{figure}

For numerical evaluation of the solution, in practice we impose the boundary condition
$q_0(\mu)=\tilde\varphi(\mu)$, etc., at a sufficiently small positive $\mu\sim 10^{-10}$. The apparent stiffness in the
first two nonuniversal equations of Eq.~(\ref{longeqsB}) at $s=t$ can be circumvented by adding to $t$ a tiny
imaginary part $\epsilon$ of the order of $O(10^{-10})$. With appropriately chosen values of $\epsilon=\Im m(t)$,
the real parts of $q_0(s)$ and $p_0(s)$ are stable upon varying $\epsilon$, and
at $s=\Re e(t)$ they vanish up to the accuracy of $O(\epsilon)$.
The joint distribution of the smallest eigenvalue $t$ and the second smallest eigenvalue $s$
of random positive-definite Hermitian matrices,
\begin{align}
{P}_{12}(t,s)=
-\Theta(s-t)
\partial_s \(
\rho_1(t)
\mathrm{Det}(\mathbb{I}-\tilde{\mathbf{K}}_{(0,s)})\)
=\Theta(s-t)
\rho_1(t) R(s)\exp\(-\int_0^s ds'R(s')\)
\label{P12Bessel}
\end{align}
obtained by this prescription for $\nu=0$ and $1$, and the corresponding two-point correlation function 
$\rho_2(t,s)=\rho_1(t)\rho_1(s)-K(t,s)^2$ for $t<s$
are converted to those of the singular values of random complex matrices by the replacements
$t \mapsto t^2, s\mapsto s^2$ and ${P}_{12}\mapsto 4ts\,{P}_{12}, \rho_{2}\mapsto 4ts\,\rho_{2}$
and are plotted in Fig.~\ref{fig:P12BesselJD}.
Again we have confirmed that the Fredholm determinant 
$\mathrm{Det}(\mathbb{I}-\tilde{\mathbf{K}}_{(\mu,s)})=\exp\(-\int_\mu^s ds' R(s')\)$
obtained by the TW system is in perfect agreement with numerical values from
the Nystr\"{o}m-type quadrature approximation.
Specifically, for a cutoff value $\mu=10^{-12}$,
relative deviations between $\int^s_\mu ds'R(s')$, computed from the TW
system (\ref{longeqsB})--(\ref{RsBessel}) using Mathematica with $\epsilon=10^{-10}$ and at quadruple {\tt WorkingPrecision},
and $-\log\mathrm{Det}(\mathbb{I}-\tilde{\mathbf{K}}_{(0,s)})$ 
computed by the Nystr\"{o}m-type approximation with quadrature
order $M=200$, are between $10^{-12}$ and $10^{-9}$ for a range of (original) variables $0\leq s,t\leq 81$
(see Table 2 for $\nu=0$, $t=4$ and $s=1,\ldots,13$). The table of numerical data for the Fredholm determinant 
$\mathrm{Det}(\mathbb{I}-\tilde{\mathbf{K}}_{(0,s)})$
(after replacements $t \mapsto t^2, s\mapsto s^2$)
is attached as online supplementary material.
\begin{table}[h]
\centering
\begin{tabular}{ccccccc}
\hline
{\small $s$} & {\small $1$} & {\small $2$} & {\small $3$} & {\small $4$} & {\small $5$} & {\small $6$}\\
{\small Rel. dev.} & 
{\small $\!\!-3.62\times 10^{-12}\!\!$} & {\small $\!\!-2.79\times 10^{-12}\!\!$} & 
{\small $\!\!-1.38\times 10^{-12}\!\!$} & {\small $\!\!-6.66\times 10^{-13}\!\!$} &
{\small $\!\!-8.32\times 10^{-10}
\!\!$} & {\small $\!\!
-1.05\times 10^{-9}
\!\!$} \\
\hline
{\small $7$} & {\small 8} & {\small 9} & {\small 10} & {\small 11} & {\small 12} & {\small 13} \\
{\small $\!\!8.04\times 10^{-10}\!\!$} & {\small $\!\!-3.01\times 10^{-10}\!\!$} &
{\small $\!\!4.24\times 10^{-10}\!\!$} & {\small $\!\!-1.19\times 10^{-9}\!\!$} &
{\small $\!\!-2.57\times 10^{-10}\!\!$} & {\small $\!\!-1.85\times 10^{-9}\!\!$} & 
{\small $\!\!-2.42\times 10^{-10}\!\!$}\\
\hline
\end{tabular}
\caption{Relative deviations of $\log\mathrm{Det}(\mathbb{I}-\tilde{\mathbf{K}}_{(10^{-12},s)})$ for the Bessel kernel
($\nu=0$)
at $t=4$ and various values of $s$
computed by the TW method ($\Im m(t)=10^{-10}$, 
{\tt WorkingPrecision $\mathtt{\to}$ 4 MachinePrecision}) 
versus the ones by the Nystr\"{o}m-type approximation ($M=200$).}
\end{table}

\section{Conclusion and perspectives}
In this article we have shown that the TW method is applicable to the evaluation of
J\'{a}nossy densities and joint eigenvalue distributions for a kernel 
$\mathbf{K}\doteq (\varphi(x)\psi(y)-\psi(x)\varphi(y))/(x-y)$
if it is applicable to the gap probability.
Essential to the inheritance of the TW criteria from
$\mathbf{K}$ to the transformed kernel 
$\tilde{\mathbf{K}}\doteq (\tilde{\varphi}(x)\tilde{\psi}(y)-\tilde{\psi}(x)\tilde{\varphi}(y))/(x-y)$
is the structure that the map between the component functions
$(\varphi, \psi)\mapsto (\tilde{\varphi}, \tilde{\psi})$
is an $\mathrm{SL}(2,\mathbb{R})$ gauge transformation to a covariantly constant section
of an $\mathbb{R}^2$-bundle with an $\mathfrak{sl}(2,\mathbb{R})$ connection
$\mathcal{A}(x)=\frac{1}{m(x)}
{\small 
\begin{bmatrix}
~~A(x) \!\!\!&\!\!\!~~B(x)\\
-C(x)\!\!\!&\!\!\!-A(x)
\end{bmatrix}} .$
Our formulation generalizes the spirit of Ref.~\cite{Forrester:1996},
which computed a special case of J\'{a}nossy  density for the sine kernel by the TW method.
As the simplest examples we evaluated the joint distributions of the two extremal eigenvalues
$P_{12}(t,s)$ for the Airy and Bessel kernels by the TW method 
and by the quadrature approximation as well, and confirmed their agreement to very high accuracies.
These results (Figs.~\ref{fig:P12AiryJD} and \ref{fig:P12BesselJD})
precisely fit the measured histograms from Gaussian-randomly generated matrices (Fig.~\ref{histograms}).

We list the pros and cons of our approach.
In contrast to the model-specific approaches in the preceeding works
\cite{Forrester:2007,Witte:2013,Perret:2014,Forrester:1996}, which computed only
the first-order J\'{a}nossy densities for the Bessel, Airy, and sine kernels,
our method is universally and systematically applicable to any kernel satisfying the TW criteria,
including but not limited to $q$-orthogonal, beyond-Airy, and various finite-$N$ kernels,
and to any $k$th-order J\'{a}nossy densities. 
In exchange, the intrinsic connection between our formulation and the
isomonodromic systems associated with Painlev\'{e} transcendents and integrability in these works
is completely obscured.
Our approach is not well suited for asymptotic analysis for $|t-s|\gg 1$ or $|t-s|\ll 1$, either.

Finally we comment on possible extensions and physical applications of our approach.
\begin{itemize}
\item
The joint distribution $P_{1\cdots k}(s_1,\ldots,s_{k})$ of the first $k$ extremal eigenvalues 
is trivially obtained by repeating the procedure (\ref{procedure}) $(k-1)$ times,
which increases the order of the polynomials $\tilde{A}(x)$, etc.~by $2(k-1)$.
The joint distribution $P_{p_1\cdots p_\ell}(s_{p_1},\ldots,s_{p_\ell})$ of 
the $p_1\mathrm{th},\ldots, p_\ell$th extremal eigenvalues follows from  $P_{1\cdots k}(s_1,\ldots,s_{k})$
by integrating out $k-\ell$ eigenvalues in an ordered cell, such as
$P_{13}(s_1,s_3)=\int_{s_1}^{s_3}ds_2\,P_{123}(s_1,s_2,s_3)$.
\item
For the applicability of the TW method to inherit from $\mathbf{K}$ to $\tilde{\mathbf{K}}$,
the requirement of the Chrsitoffel--Darboux form (\ref{CD}), 
characteristic of $\mathrm{U}(N)$ invariant ensembles, can actually be relaxed to 
more generic, asymmetric kernels of
the integrable class \cite{Its:1990}:
\begin{align}
K(x,y)=\sum_{\ell=1}^r\frac{f_\ell(x) g_\ell(y)}{x-y}:=\frac{\bs{f}(x)\cdot \bs{g}(y)}{x-y}
~~,~~
\bs{f}(x)\cdot\bs{g}(x)=0.
\label{Kr}
\end{align}
Here $r$-component real functions 
$\bs{f}(x)=\(f_1(x),\ldots,f_r(x)\)^t$ and $\bs{g}(y)=\(g_1(y),\ldots,g_r(y)\)^t$
are covariantly constant sections for some meromorphic $\mathfrak{sl}(r,\mathbb{R})$ connections
$\mathcal{A}(x)$ and $\mathcal{B}(y)$,
respectively.
In this generalized case, 
an $\mathrm{SL}(r,\mathbb{R})$ gauge transformation on them,
\begin{align}
\begin{array}{ll}
{\dps \bs{f}(x)\mapsto
\tilde{\bs{f}}(x)=\bs{f}(x)-\frac{K(x,t)}{K(t,t)}\bs{f}(t)=\mathcal{U}(x)\bs{f}(x)} \\
{\dps \bs{g}(y)\mapsto
\tilde{\bs{g}}(y)=\bs{g}(y)-\frac{K(t,y)}{K(t,t)}\bs{g}(t)=\mathcal{U}(y)^{-1\,t}\bs{g}(y)}
\end{array} ,~~
\mathcal{U}(x)=\mathbb{I}-\frac{\bs{f}(t)\bs{g}(t)^t}{\rho_1(t)(x-t)}
\end{align}
maps $K(x,y)$ to $\tilde{K}(x,y)=\tilde{\bs{f}}(x)\cdot\tilde{\bs{g}}(y)/(x-y)$
while retaining $\tilde{\bs{f}}(x)\cdot\tilde{\bs{g}}(x)=0.$
An example of a kernel of type (\ref{Kr}) is the Pearcey kernel (with $r=3$) governing
spectral correlations of random matrices in an external source,
$H=H_{\mathrm{GUE}}+c\,\mathrm{diag}(\mathbb{I}_{N/2}, -\mathbb{I}_{N/2})$
in the critical regime where a gap in the eigenvalue support closes at the origin \cite{Brezin:1998}.
This ensemble schematically models the QCD Dirac operator at finite temperature
\cite{Stephanov:1996}.
Application of our strategy to its J\'{a}nossy density 
by the generalized TW method \cite{Tracy:2006} will be reported in a separate publication.
\item
Ensembles of Dirichlet 
$L$-functions are acknowledged as ideal quantum-chaotic systems for
their distributions of zeroes on the critical line \cite{Katz:1999,Keating:2000}.
It is well anticipated but worth verifying
that the joint distributions of the two smallest zeroes of
$L$-functions are described by J\'{a}nossy densities for the Bessel kernels (\ref{P12Bessel})
at $\nu=\pm1/2$, depending on the sign in the functional equation
of the $L$-functions.
\item
In the context of noncritical string theory, 
conditioning the loci of some ($k$) of $N$ eigenvalues of matrix models at (multi)criticality
(i.e.,~beyond Airy)
outside their main support has been interpreted as introducing $k$ ZZ branes to the Liouville theory
\cite{Hanada:2004,Sato:2005}.
As all efforts have been concentrated on extracting {\em leading} nonperturbative corrections to the free energy
in the large-$N$ limit, it is worthwhile to apply our analytic strategy for computing
the J\'{a}nossy density $J_k(\{x\};I)$ to those models and obtain unapproximated, 
fully nonperturbative free energy that incorporates {\em all} D-brane contributions.
\item
It seems less promising to extend our strategy to J\'{a}nossy densities of 
quaternion kernels \cite{Soshnikov:2003} governing orthogonal and symplectic ensembles \cite{Tracy:1996}, 
or transitive ensembles interpolating different symmetry classes.
Nevertheless, the observation that the J\'{a}nossy density for these cases 
is expressed as a Fredholm Pfaffian of the transformed quaternion kernel
\cite{Nagao:2000,Nagao:2001},
$\mathrm{Det}\(\mathbb{I}-(\mathbf{K}-\bs{k}^t {\boldsymbol\kappa}^{-1} \bs{k})_I\)^{1/2}$,
always permits numerical evaluation by the quadrature approximation.
Currently we are exploring 
the application of this strategy to
the quaternion kernel of the chGSE-chGUE transitive ensemble \cite{Forrester:1999},
to obtain individual distributions of the staggered Dirac operator of two-color QCD 
at finite baryon-number chemical potential $\mu$ and {\em with dynamical quarks of masses} $m_f$
introduced as the conditioned eigenvalues $x_f=-m_f^2$,
extending our previous work \cite{Yamamoto:2018} on the quenched case.
\end{itemize}

\section*{Acknowledgments}
I thank Peter Forrester for helpful comments on the manuscript.
This work is supported in part by a JSPS Grant-in-Aid for Scientific Research (C) No.~7K05416.

\section*{Supplementary material}
Numerical data of the Fredholm determinant 
$\mathrm{Det}(\mathbb{I}-\tilde{\mathbf{K}}_{(s,\infty)})$
for the Airy kernel in the range $-7\leq s,t \leq 5$
are attached as {\tt JanossyAiry.nb}.
Numerical data of the Fredholm determinant 
$\mathrm{Det}(\mathbb{I}-\tilde{\mathbf{K}}_{(0,s)})$
(after replacements $t \mapsto t^2, s\mapsto s^2$)
for the Bessel kernels at $\nu=0$ and $\nu=1$ in the range $0\leq t,s \leq 9$
are attached as {\tt JanossyBessel.nb}.

\end{document}